\documentclass[12pt,preprint]{aastex}

\newcommand{\irassrc}{{\it IRAS} 06337+1051}

\newcommand{\IRAS}{{\it IRAS}}

\newcommand{\MAP}{{\it MAP}}

\newcommand{\LPH}{LPH~201.663+1.643}

\newcommand{\HII}{H\,{\scriptsize II}}

\newcommand{\MJysr}{MJy~sr$^{-1}$}
\newcommand{\etal}{{\it et al.}~}

\shorttitle{\LPH}
\shortauthors{McCullough and Chen}

\begin{document}

\title{An Alternative to Spinning Dust for the Microwave Emission of \LPH:
an Ultracompact \HII\ Region}

\author{P. R. McCullough\altaffilmark{1} and R. R. Chen\altaffilmark{}}
\affil{Dept. of Astronomy, Univ. of Illinois, Urbana IL 61801, USA}
\altaffiltext{1}{Cottrell Scholar of Research Corporation.}
\email{pmcc@astro.uiuc.edu, raychen@astro.uiuc.edu}

\begin{abstract}

The microwave spectral energy distribution of the dusty, diffuse \HII\ region
\LPH\ has been interpreted by others as tentative evidence for
microwave emission from spinning dust grains. We present an alternative
interpretation for that particular object; specifically, that an ultracompact
\HII\ region embedded within the dust cloud would explain the available 
observations as well or better than spinning dust.
Parameters for the size, surface brightness, and flux density of the putative
ultracompact \HII~region,
derived from the microwave observations, are within known ranges.
A possible candidate for such an ultracompact \HII\ region is
\irassrc, based upon its infrared colors. However, \irassrc's infrared flux
appears to be too small to be consistent with the microwave flux required
for this alternative model to explain the observations.

\end{abstract}


\section{Introduction} \label{sec_intro}

Studies of the cosmic microwave background are affected by foreground
emissions from the Milky Way and other sources. Principal mechanisms
are thermal emission from warm dust, synchrotron from electrons gyrating
in magnetic fields, and free-free (a.k.a. bremsstrahlung) emission from
ionized plasma. An elusive forth foreground is attributed to spinning dust
\citep[]{draine98a} or perhaps magnetic dust \citep[]{draine99}.
A small but growing number of statistical comparisons
of infrared tracers of dust and microwave observations of ``CMB quality''
have demonstrated repeatedly, but in each case at low significance,
an excess of microwave emission, correlated with dust on the sky,
and possibly due to spinning or magnetized dust. 
Finkbeiner {\it et al.} (2002, Paper I) 
list such statistical comparisons and add to them two tentative
detections of the spinning-dust mechanism in specific astronomical sources,
\LPH\ and Lynds 1622, from a sample of 10 sources observed by them.

\LPH\ is the topic of this {\it Letter}. It differs from the other nine dust
clouds observed by Finkbeiner {\it et al.} (2002) in that it is known to be
a diffuse \HII\ region whereas the others are not ionized.
Finkbeiner {\it et al.}
included \LPH\ in their sample despite their prejudice that free-free
emission might overwhelm any emission from spinning dust. 
Alternatively, one might anticipate that ionized regions would be good
places to find microwave emission from spinning dust, because in those regions,
ion collisions with grains are expected to be
the largest contributory factor by far in maintaining the spin of the grains
\citep[]{draine98b}.

Because of its special status as the sole \HII\ region in the list, and
because \LPH\ provided the only very significant detection and was indeed
much brighter than the spinning-dust theory would have predicted, we 
thought that a different explanation might exist.
The alternative model for \LPH\ that we present in this {\it Letter}
is the superposition of one source with very large emission measure
and very small angular scale with a low emission-measure, extended source.
The combination of a small, optically thick source and a large optically thin
source, both emitting by free-free, can create the rising microwave spectrum
observed in Paper I.

\section{Observations} \label{sec_obs}

The three observations from Table 3 of Paper I relevant to this {\it Letter} are
$\rm I_\nu = 0.397 \pm 0.046,\ 0.516 \pm 0.019,\ and\ 0.667 \pm 0.027$\ 
\MJysr\ at $\nu =$ 5 GHz, 8.25 GHz, and 9.75 GHz. Note that the brightness
increases with frequency.

The envelopes of newly forming stars in close proximity to an O-type star
can have emission measures as large as $\rm 3\times 10^8~cm^{-6}~pc$
over angular scales $^{<}_{\sim} 1$\arcsec\ (McCullough \etal\ 1995).
An ensemble of externally ionized stellar envelopes
could match the angular extent of the dust cloud, but the combined flux
from the ensemble would be insufficient to match the
observed microwave flux from \LPH\, which at 9.75 GHz is $\sim 3$ Jy.
For example, the total 15 GHz flux density from the the ensemble of 24 ionized
envelopes illuminated by $\theta^1$C Orion is 0.175 Jy (Felli \etal\ 1993).
A better idea would be an 
ultracompact \HII\ region, which would intercept all of the UV flux
from an OB star, whereas an ensemble of stellar envelopes intercepts only a tiny
fraction of the OB star's flux.

In order to estimate the distance to \LPH\,
we searched its vicinity in the SIMBAD database for hot stars
that could ionize it and found two,
GSC 00737-01170 (O5:; V=11.29; B-V=0.75) and GSC 00737-00898 (O5; V=10.27;
B-V=1.11), where the spectral types and magnitudes are from 
Voroshilov \etal\ (1985).
For an O5 V star, $\rm M_V = -5.3$ and for an O5 Ia star, $\rm M_V = -6.4$
(Vacca, Garmany, \& Shull 1996). Regardless of luminosity class, an unreddened
O5 star has B-V = $-0.33$ (Cox 2000). 
For $A_V = 3.2 E(B-V)$, the implied distances to 
GSC 00737-01170 and GSC 00737-00898 are 4.2 kpc and 1.6 kpc, respectively
if each is O5 V, and 7.0 kpc and 2.6 kpc, respectively
if each is O5 Ia. At $l=201.6$\arcdeg\ and $b=+1.6$\arcdeg, \LPH\ is
$\sim$22\arcdeg\ from the anticenter direction, and its Galactic height
$Z < 200$ pc for d $< 7.0$ kpc.

\section{Modeling} \label{sec_modeling}

The optical depth for free-free emission,
\begin{equation}
\rm \tau_{ff} \approx 2.6~T_4^{-1.35}~(EM/10^9~cm^{-6}~pc)~(\nu/10~GHz)^{-2.1},
\label{eq-tauff}
\end{equation}
where
$\rm T_4 \equiv T_e / 10^4 K$,
$\rm T_e$ is the electron temperature,
$\rm EM$ is the emission measure, and
$\nu$ is the radio frequency.
As observed with an antenna of beam solid angle $\Omega$,
the apparent microwave surface brightness 
of a uniform disk of radius $\theta_s$ is
\begin{equation}
\rm I_1 = B_\nu(T_e) ( 1 - e^{-\tau_{ff}} ) \pi \theta_s^2 \Omega^{-1}
\label{eq-i1}
\end{equation}
where $\rm B_\nu(T_e)$ is the Planck function and we assume that
the background is negligible and that $\pi \theta_s^2 << \Omega$.
For the 140-foot antenna, $\Omega = 4.59, 1.75,$ and 1.31 
microsteradians for observing frequencies of 5 GHz, 8.25 GHz, and 9.75 GHz
respectively, but in our model we use $\Omega = 4.59\times 10^{-6}$ sr
for all frequencies because in Paper I the higher frequency data were smoothed
to match the 5 GHz beam.

To get the total emission of the model, we add $\rm I_1$ to the
free-free emission from a diffuse, optically thin component.
The diffuse component cannot be very diffuse, because
if it was larger than the 0.2\arcdeg\ chop used in Paper I
at 8.25 and 9.75 GHz, it would be 
eliminated by the chopping. Although we do not know the angular distribution
of free-free emission in this region, we assume that the diffuse free-free
component contributes to the brightnesses at each
frequency equally except for the $\nu^{-0.1}$ spectral index of
optically thin free-free. Thus, our model is
\begin{equation}
\rm I_\nu = I_2 (\nu/5~GHz)^{-0.1} + B_\nu(T_e) ( 1 - e^{-\tau_{ff}} ) \pi \theta_s^2 \Omega^{-1},
\label{eq-inu}
\end{equation}
where the model parameters are $\rm I_2$
and the two parameters of the ultracompact \HII\ region (its radius
$\theta_s$ and its emission measure, EM).

Fig. \ref{fig_spectrum} compares the observed brightnesses
to those predicted by the model
with a few combinations of parameters for illustration.
Fig. \ref{fig_contours} illustrates $\chi^2$ contours in the phase space of the parameters
of the ultracompact \HII\ region, $\rm \theta_s\ and\ EM$,
with $\rm I_2$ fixed at 0.255 \MJysr.
The allowed parameters are similar to those of known
ultracompact \HII\ regions (Wood \& Churchwell 1989b).
For $\rm EM = 10^9$ cm$^{-6}$~pc, $\theta_s = 1$\arcsec, 
the number of ionizing photons required to power the ultracompact
\HII\ region,
$N_c^\prime~^{>}_{\sim} 4\times10^{47} (d/1~{\rm kpc})^2 {\rm s}^{-1}$
(Kurtz, Churchwell, \& Wood 1994),
which corresponds to a main-sequence spectral type B0 or earlier
for $d = 1.6$ kpc
or O8 or earlier for $d = 4.2$ kpc (Vacca, Garmany, \& Shull 1996).

\section{A Possible Candidate: \irassrc} \label{sec_candidate}

\irassrc\ stands out in the \IRAS\ ISSA images as the brightest 25 $\mu$m
source in the vicinity of \LPH.
\irassrc\ has the infrared colors of an ultracompact \HII\ region (Fig.
\ref{fig_wood1b}).
Its PSC position (Beichman \etal\ 1988),
$\rm 6^h36^m29^s.47~+10\arcdeg49\arcmin5\arcsec.1$ [2000],
with an error ellipse $38\arcsec \times 7\arcsec$\ 
oriented with the major axis at a position angle of 94\arcdeg,
is compatible with the Paper I's linear scans, which were
centered on $\rm 6^h36^m40^s~+10\arcdeg46\arcmin28\arcsec$
and oriented at position angle 292.5\arcdeg.
The nominal scan's closest approach to the PSC position of \irassrc\ 
is 1.4\arcmin,
and the peak along the nominal 8.25 GHz scan occurs 2.4\arcmin\ or
0.6\arcmin\ east of the nominal position of \irassrc, for the forward
and reverse scan directions respectively (Paper I, Fig. 3b).
For comparison, the resolution (FWHM) of the microwave data was 6\arcmin.
The antenna temperature at 8.25 GHz of the
ultracompact \HII\ region is $^{<}_{\sim} 55$\% of the total for the
models in Section \ref{sec_modeling},
so the peak position at 8.25 GHz need not be centered
at the position of the ultracompact \HII\ region.

A potential flaw in \irassrc\ as the putative ultracompact \HII\ region
is that our model requires parameters for the
ultracompact \HII\ region that would make it brighter than most ultracompact
\HII\ regions at radio wavelengths (cf. Wood \& Churchwell 1989a and
Kurtz, Churchwell, \& Wood 1994) but at $60\mu$m
\irassrc\ is four times fainter than
the median of embedded OB star candidates selected by Wood \& Churchwell
(1989a). 
For the models depicted in
Fig. \ref{fig_contours}, the ultracompact \HII\ regions' flux density,
$F_\nu(2 cm)$ ranges from 1.6 Jy to 4.4 Jy.
Ultracompact \HII\ regions are more than a thousand times brighter 
in the infrared than in the radio (Kurtz, Churchwell, \& Wood 1994),
so the $3\sigma$ upper limit to the \IRAS\ PSC flux at 100 $\mu$m,
$F_\nu({\rm 100\mu m}) < 383$ Jy would imply that the radio flux density 
$F_\nu(2 cm) < 0.4 $ Jy, i.e. in conflict with the model fluxes.
Stated another way, the luminosity is 
\begin{equation}
L_* = 4 \pi d^2 S_\nu \Delta\nu f_\nu^{-1},
\label{eq-lum}
\end{equation}
where $f_\nu$ is the fraction of total luminosity $L_*$ emitted
at frequency $\nu$ in the passband $\Delta\nu$, and $S_\nu$ is
the associated flux density.
For \irassrc\ at 60 $\mu$m, $S_\nu = 33$ Jy,
$\Delta\nu = 2.58\times10^{12}$ Hz (Beichman et al. 1988),
and $f_{\rm 60 \mu m} = 0.25$ for ultracompact \HII\ regions
(Wood \& Churchwell 1989a), so $L_* = 100 (d /1 kpc)^2 L_\odot$.
Because ionizing OB stars have $L_* > 10^4 L_\odot$,
the implied distance $d > 10$ kpc.
At such large distances from the Galactic center,
massive stars do exist but are rare (de Geus \etal\ 1993).
However, if the distance to \irassrc\ $d > 10$ kpc, then it is too
far behind \LPH\ for them to be 
physically associated, and their alignment would be highly improbable
{\it a priori}.

\section{Conclusions} \label{sec_dis} 

An optically thick ultracompact \HII\ region can provide a
rising microwave spectrum
with a spectral index $\approx 2$ which is similar at these
frequencies to the spinning-dust model's spectral index $\approx 2.8$
(Ferrara \& Dettmar 1994; Draine \& Lazarian 1998b).
For free-free emission,
for $\tau_{ff} \ ^{>}_{\sim} 1$ at $\nu ^{<}_{\sim} 10$ GHz, the 
EM must be $^{>}_{\sim} 10^9$ cm$^{-6}$~pc. 

While \irassrc\ is within a fraction of a 6\arcmin\ beam width of \LPH,
and has the infrared colors
of an ultracompact \HII\ region, its \IRAS\ fluxes are too small unless
its distance is improbably large. Also, the distance-independent ratio of
the 100 $\mu$m flux of \irassrc\ to our model's 2-cm flux is too small by
more than an order of magnitude.
The latter conflict is mitigated somewhat if the star is an early B-type star
that has an atypically large ratio of ionizing to bolometric luminosity
as has been observed in some cases (Cassinelli 1996) and has been predicted by
modern stellar atmosphere models (Vacca, Garmany, \& Shull 1996).

Paper I concluded that ``as long as the density is low,
as in the LPH list, \HII\ regions may be optimal targets for future work for
DASI, CBI, and \MAP.'' 
It would be inappropriate to misinterpret this single case study of \LPH\ as
contradictory to that recommendation. 
Instead, attention to its premise, that the density be low,
is especially appropriate. The all-sky observations made by \MAP\ will allow
astronomers to conveniently avoid any regions that might be unsatisfactory
in that regard.

The hypothesis presented in this {\it Letter} can be tested directly with radio
interferometric imaging, which we plan to complete in early 2002.
If that test shows not even one dense ionized knot in 
\LPH, then this {\it Letter}'s remaining value
will be the analysis of Section \ref{sec_modeling}, and the
microwave emission from \LPH\ will still be a mystery.

\acknowledgments

We thank Doug Finkbeiner for sending us a pre-publication manuscript
that initiated our thoughts about \LPH. A conversation with Bruce Draine
during his visit to Illinois on October 30, 2001, was especially
helpful; as referee, B. D. made additional valuable comments,
most importantly that \irassrc\ be in its own section.

P. R. M. wishes to acknowledge two grants that funded this research,
a CAREER AWARD from the National Science Foundation (AST-9874670) and a Cottrell
Scholarship from the Research Corporation,
and to thank the persons that championed the associated proposals.

\clearpage

\figcaption{\label{fig_spectrum}
The brightness of \LPH\ at three microwave frequencies (Paper I)
is compared to that of combined emission from a diffuse \HII\ region
and an ultracompact \HII\ region (Equation \ref{eq-inu}) for various
model parameters (solid lines). For models A, B, C, and D respectively, with
$\rm \theta_s\ and\ EM$\ as indicated in Fig. \protect{\ref{fig_contours}},
and $\rm I_2 = 0.255$ \MJysr\ in all four models,
the $\rm \chi^2 \approx 2, 4, 10, and\ 20$.
The dashed line shows model A with the additional contribution from
spinning dust appropriate for a warm ionized medium (Ferrara and Dettmar 1994;
Draine \& Lazarian 1998b)
and a dust column $\rm N(H) = 2.7 \times 10^{22} cm^{-2}$ (Paper I).
}

\figcaption{\label{fig_contours}
Contours of $\chi^2$ are plotted in the phase space of the parameters
of the ultracompact \HII\ region, $\theta_s$ and $\rm EM$,
for a representative value for the parameter that describes
the diffuse \HII\ region $\rm I_2$ = 0.255 \MJysr.
For values of EM $^{>}_{\sim} 10^9$
cm$^{-6}$~pc, the ultracompact \HII\ region is optically thick at even
the highest observed frequency, so
the $\chi^2$ contours are open at the top of the plot.
Models with
$\rm \theta_s\ and\ EM$\ as indicated by the letters A, B, C, and D
are compared to the data in Fig. \protect{\ref{fig_spectrum}}.
}
 
\figcaption{\label{fig_wood1b}
\irassrc~is compared to known ultracompact \HII~regions ({\it filled circles})
in an \IRAS\ color-color plot from Wood \& Churchwell (1989a).
Two solid-lined rectangles are centered on the most-likely colors of
\irassrc,
${\rm Log(F_\nu(25 \mu m) / F_\nu(12 \mu m)) = 0.96 \pm 0.049}$, and
${\rm Log(F_\nu(60 \mu m) / F_\nu(12 \mu m)) = 1.59 \pm 0.094}$.
The smaller rectangle has a $1\sigma$ half-width and half-height;
the other rectangle is three times larger.
That the rectangles lie in the region of the diagram populated by
ultracompact \HII~regions and that other types of sources very rarely
populate that region make \irassrc\ likely to be an ultracompact \HII\ region.
The dashed lines indicate
the boundary prescribed by Wood \& Churchwell (1989a) to discriminate
embedded OB stars from other sources in the \IRAS\ point source catalog.
Other representative sources are those within a $2\arcdeg \times 2\arcdeg$\ 
box in the Galactic plane ({\it open squares}) and a group of \IRAS\ sources
that lie between ${\rm 13^h00^m\ and\ 13^h10^m}$ in right ascension
({\it crosses}).
}

\begin{figure}
\plotone{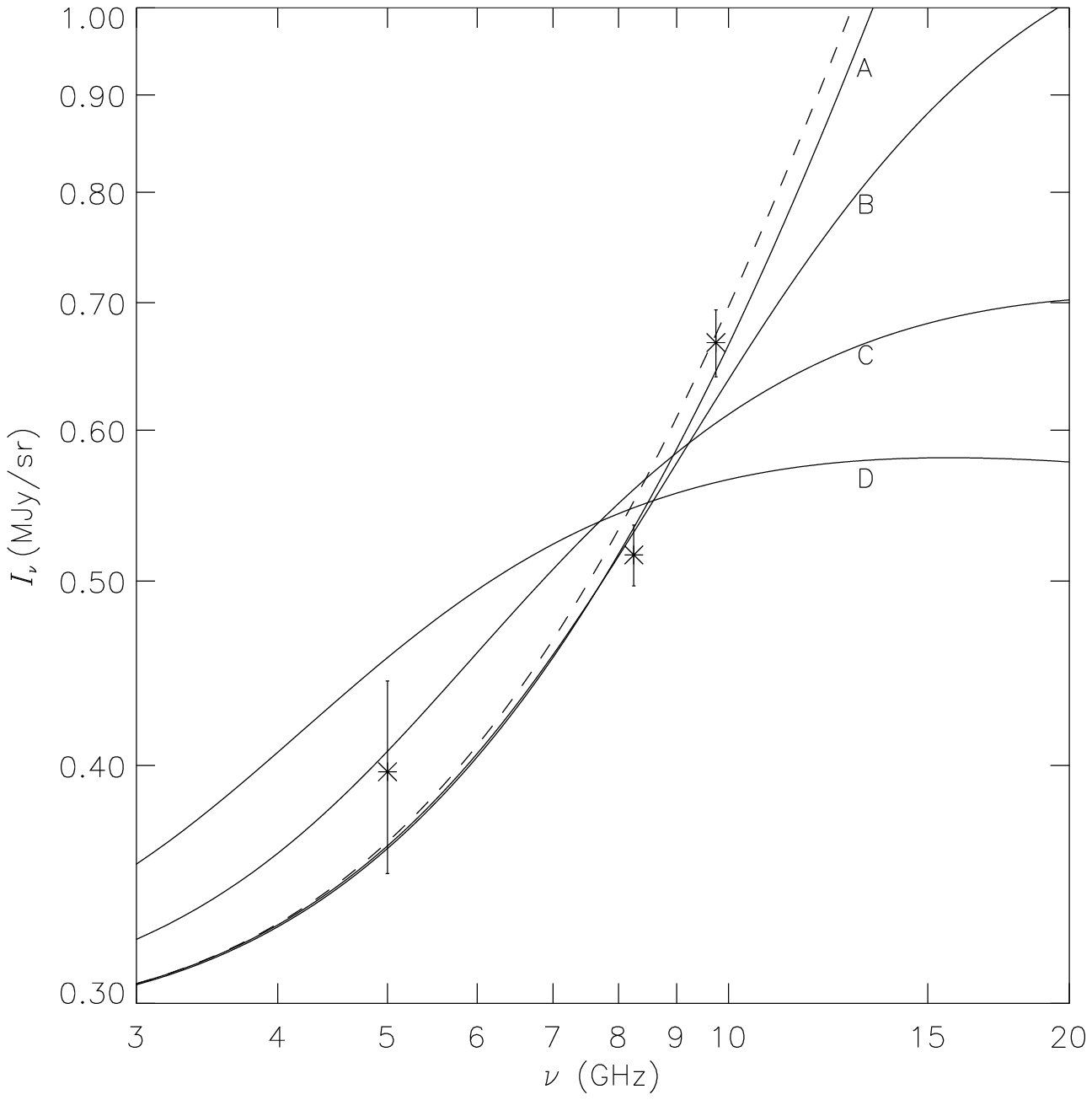}
\end{figure}

\begin{figure}
\plotone{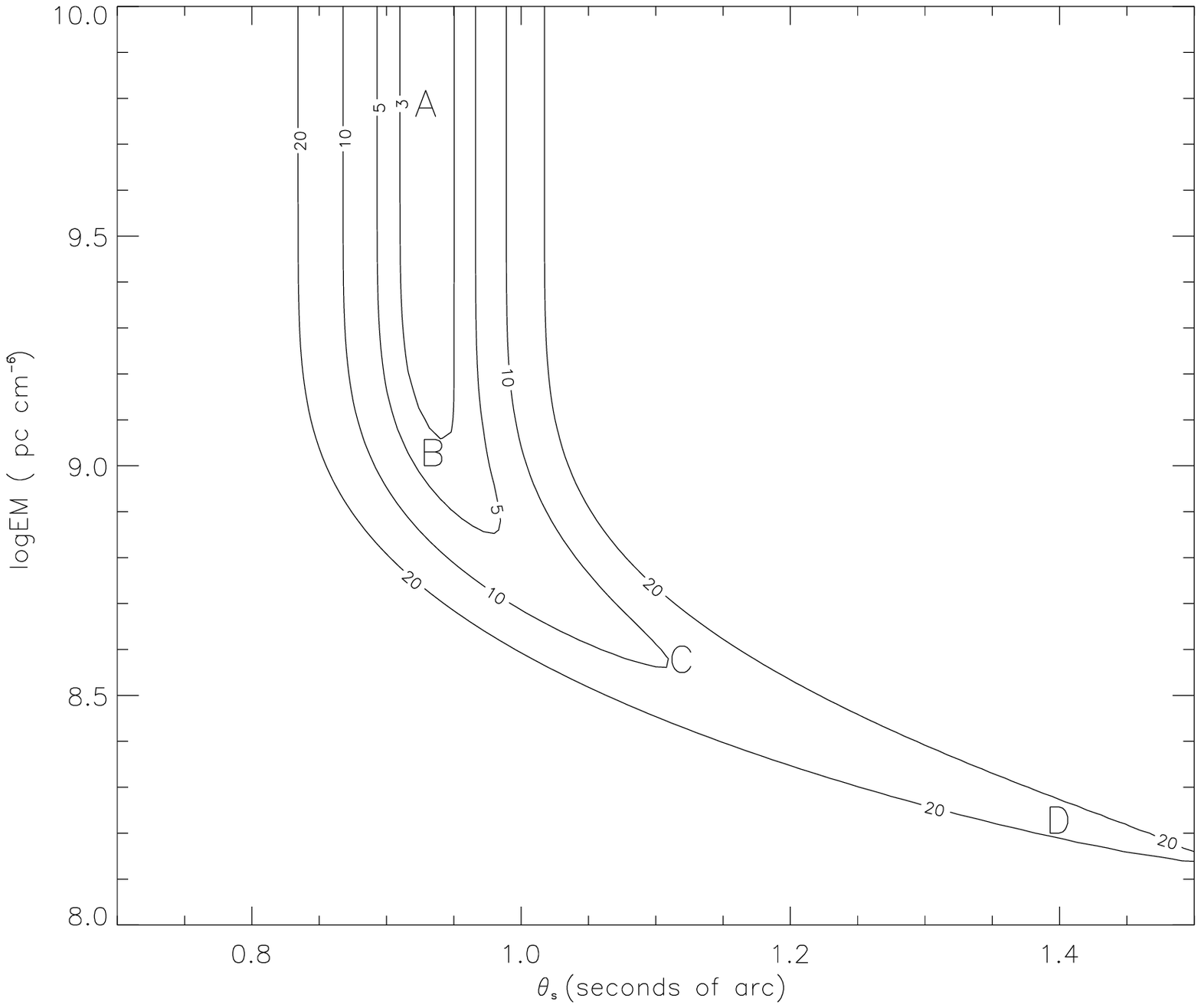}
\end{figure}

\begin{figure}
\plotone{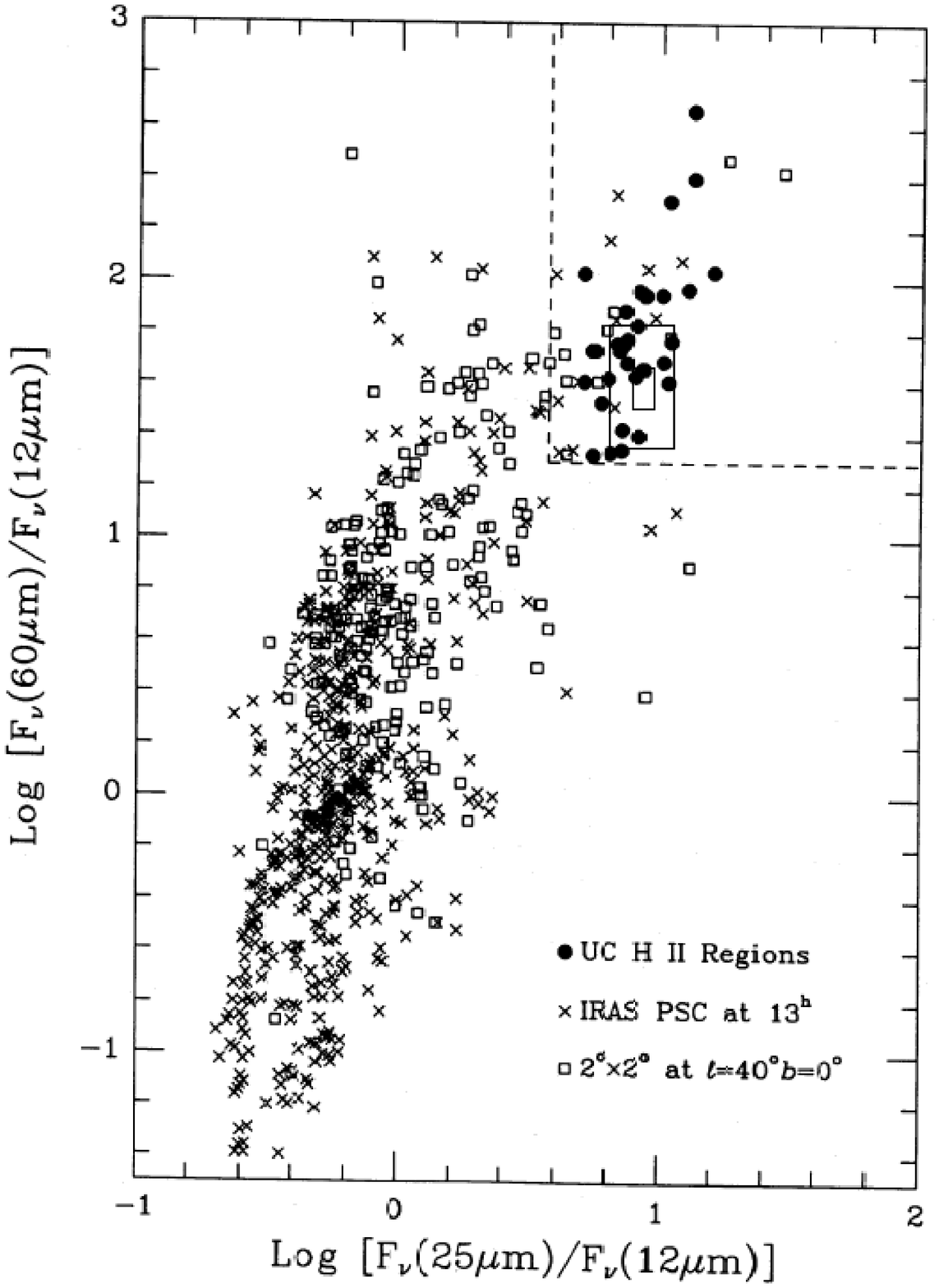}
\end{figure}

\end{document}